\documentclass[journal,letterpaper]{IEEEtran_nssmic}

\usepackage{cite}
\usepackage{graphicx}
\usepackage{subfigure}

\newcommand{\degree}{~$^{\circ}\mbox{C}\;$}

\newcommand{\co}{$^{57}$Co~}

\begin{document}
%
\title{A Prototype Si/CdTe Compton Camera and the Polarization Measurement}

\author{%
Takefumi Mitani, Takaaki Tanaka, Kazuhiro Nakazawa, Tadayuki Takahashi,
Takeshi Takashima,\\ 
Hiroyasu Tajima, Hidehito Nakamura, Masaharu Nomachi, 
Tatsuya Nakamoto, Yasushi Fukazawa%
\thanks{Manuscript received November 15, 2003;
revised April 15, 2004.}%
\thanks{T. Mitani, T. Tanaka, and T. Takahashi
are with the Institute of Space and Astronautical Science (ISAS),
Japan Aerospace Exploration Agency (JAXA),
Sagamihara, Kanagawa 229-8510, Japan,
and also with the Department of Physics, the University of Tokyo,
Bunkyo, Tokyo 113-0033, Japan
 (telephone: 81-42-759-8510, e-mail:mitani@astro.isas.jaxa.jp).}
\thanks{K. Nakazawa and T. Takashima are with the Institute of Space
and Astronautical Science (ISAS),
Japan Aerospace Exploration Agency (JAXA),
Sagamihara, Kanagawa 229-8510, Japan.}
\thanks{H. Tajima is with Stanford Linear Accelerator Center,
Menlo Park, CA 94025 USA.}
 \thanks{
H. Nakamura and M. Nomachi are with the Laboratory of Nuclear Studies,
Osaka University, Toyonaka, Osaka 560-0043, Japan}
 \thanks{
T. Nakamoto, Y. Fukazawa are with  Department of Physics, 
Hiroshima University, Higashi Hiroshima, Hiroshima 739-8526, Japan}
}
\maketitle

\begin{abstract}
A Compton camera is the most promising approach for gamma-ray
detection in the energy region from several hundred keV to MeV, especially
for application in high energy astrophysics. In order to
obtain good angular resolution,  semiconductor detectors
such as silicon, germanium and cadmium telluride(CdTe) have several
advantages over scintillation detectors, which have been used so far.
Based on the recent advances of high resolution CdTe
and silicon imaging detectors, we are working on a Si/CdTe Compton camera.
We have developed  64-pixel CdTe detectors with a pixel size of 2~mm$\times$2~mm
and double-sided Si strip detectors(DSSDs) with a position resolution of 800$\mu$m.
As a prototype Si/CdTe Compton camera, we use a DSSD as a scatterer
and two CdTe pixel detectors as an absorber. In order to verify its performance,
we irradiate the camera with 100\% linearly polarised 170~keV $\gamma$-rays
and demonstrate the system works properly as a Compton camera.
The resolution of the reconstructed scattering angle is 22$^\circ$(FWHM).
Measurement of polarization is also reported.
The polarimetric modulation factor is obtained to be
43\%, which is consistent with the prediction of Monte Carlo simulations.

\end{abstract}


\section{Introduction}


\PARstart{A}s gamma-ray energies increase to the MeV region, the detection of
photons, the determination of the incident direction, and the rejection of
background become very difficult. The use of the technique of the
Compton telescope seems to be the only way for the detection of MeV
$\gamma$-rays.
The development of the technique began in 1970s
in order to explore the $\gamma$-ray universe in this energy
band\cite{ref:NIM1973_Schoenfelder, ref:NIM1975_Herzo}. 
The COMPTEL instrument on-board
the CGRO satellite, which was launched in 1991, achieved pioneering results
with this technique\cite{Ref:COMPTEL}. 
In order to go beyond the level achieved by the COMPTEL, 
the next generation of Compton telescope needs
to be developed. 

In a Compton telescope,
an incident gamma-ray photon is scattered in the scattering part of the detector
and the Compton-scattered photon is absorbed  in the absorber part (see Fig.~\ref{fig:comp}(a)).
The scattering angle $\theta$ of the incident gamma-ray is then written as,
\begin{equation}
     \cos \theta = 1- \frac{m_ec^2}{E_2}+\frac{m_ec^2}{E_1+E_2} ,
     \label{eqn:comp}
\end{equation}
where $E_1$ is the energy of the recoil electron in the scattering detector
and $E_2$ is the energy deposited in the absorbing detector.
Thus, by measuring the energy and position of the individual interactions,
we can limit the direction of the incident photon to the surface of a cone.
The projection of this cone onto the celestial sphere is the ``event circle''.
The $\gamma$-ray source is located at the point of the intersection of all event circles.
As clearly seen in the equation, the angular resolution becomes better
if the energy resolution becomes higher.

Since the major goals for the next generation Compton telescope are to increase the
detection efficiency by up to two orders of magnitude over the COMPTEL and
to realize much improved background rejection, 
the development of 3-D position sensitive detectors
with a high detection efficiency, a high angular resolution and an
excellent energy resolution is the key.
Several detector options are under consideration,
including the use of
position-sensitive solid state detectors such as  
germanium and silicon\cite{ref:KamaeNIM1987,ref:KamaeIEEE1988,ref:ACT,ref:KurfessNIM2003,ref:MEGA,ref:TIGRE,ref:CZT_SPIE1999,ref:TakahashiSPIE2003,Ref:Takahashi_NeXT}.
As to applications to nuclear medicine,
see D.~Meier et al.\cite{ref:SiComptonMedical} and references therein.

We have been working on
a semiconductor Compton telescope
based on silicon and cadmium telluride (Si/CdTe Compton camera)
in view of
sub-MeV to MeV detectors with high sensitivity\cite{ref:TakahashiSPIE2003,Ref:Takahashi_NeXT}.
In this paper, we demonstrate the first results from 
a prototype Si/CdTe Compton camera.
Firstly we describe the performance of
newly-developed CdTe pixel detectors which are used as the 
absorber part of the prototype. Secondly, the reconstruction of the scattering angle
and a polarization measurement are reported based on experiments
at the SPring-8 synchrotron radiation facility in Japan.
Further report on the imaging performance of the Si/CdTe Compton
camera will be described in a separate paper\cite{ref:TanakaSPIE2004}.

\section{Si/CdTe Compton camera}

\begin{figure}
\centerline{\includegraphics{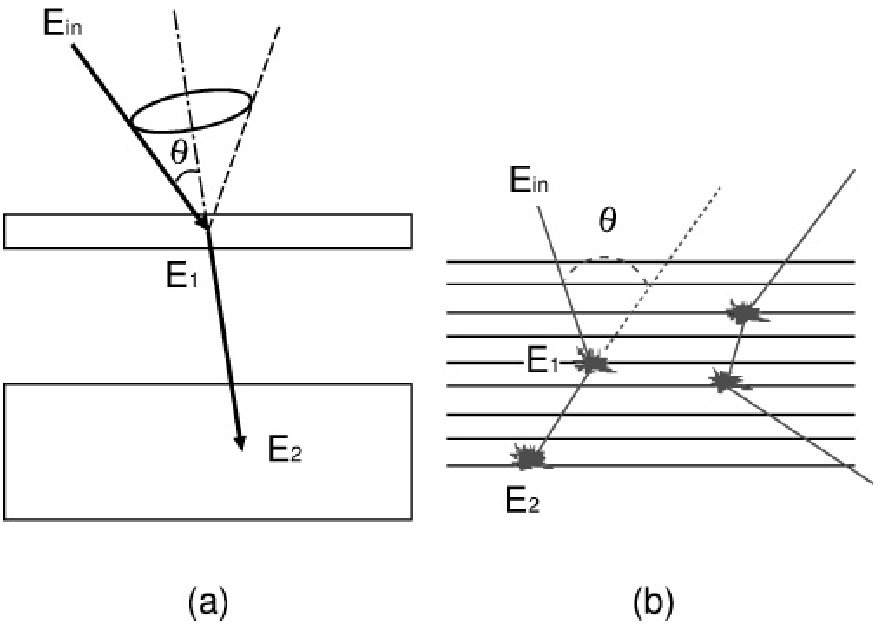}}
\caption{(a) Concept of a Compton camera. When an incident photon undergoes
a Compton scattering and the scattered photon is absorbed completely,
the scattering angle can be calculated with
Eq.~(\ref{eqn:comp}) using the energy deposited.
From the interaction position and the scattering angle, the direction the photon comes from
can be constrained. (b) Concept of a semiconductor Compton telescope.
It is a stack of position sensitive detectors made of semiconductor materials.
The upper layers work as scatterers and the lower layers work as absorbers.}
\label{fig:comp}
\end{figure}

In the Si/CdTe Compton camera, a stack of many thin scatterers made of fine pitch double-sided 
Si strip detectors is used to record the Compton scatterings. Additional layers made of a heavy
semiconductor such as CdTe (or CdZnTe) are used to absorb the Compton scattered photons.
The order of the interaction sequences, hence 
the correct energy and direction of the incident photon,  can be reconstructed by examination of the energy-momentum conservation for all possible sequences
 (Fig.~\ref{fig:comp}(b)).

For imaging devices,
their good energy resolution and the ability
to fabricate compact arrays are very attractive features
of semiconductor detectors
in comparison with inorganic scintillation detectors.
The high density and good detection efficiency of
CdTe and CdZnTe,
comparable with that of NaI(Tl) and CsI(Tl),
are important additions when we design the absorber part of the Compton telescope
\cite{ref:TakahashiSPIE2003}.
Based on our recent developments,
the CdTe diode now shows
an energy resolution
comparable with that of a Ge detector,
even at room temperature\cite{ref:IEEE2000-1, ref:TanakaNAR2003, ref:IEEE_GR}.
The energy resolution achieved for a planar diode detector,
which has dimensions of 2~mm $\times$ 2~mm as an active area and a thickness of 0.5~mm,
is 0.93~keV and 1.2~keV (FWHM) at
59.5~keV and 122~keV, respectively.

The merit of using Si as scatterers
in the energy range from 50~keV up to several hundred keV
is that the cross section of Compton scattering is well 
above that of photoelectric absorption.
Therefore the efficiency of Compton scattering is high
if the scattering part has sufficient thickness.
Additionally, Si works better than CdTe in terms of
the ``Doppler broadening'' effect.
The standard Compton formula Eq. (\ref{eqn:comp}) assumes the scattering 
takes place with a free electron at rest. 
When we take
the momenta of bound electrons into account,
we need to consider
additional uncertainties in the energy and angle of the scattered 
gamma ray\cite{ref:Doppler}. This phenomenon is often referred to as 
Doppler broadening. The additional uncertainties on the energy 
resolution and angular resolution are especially pronounced 
at low gamma ray energies and 
become less significant as the atomic number decreases.
Together with the high absorption efficiency of CdTe,
a Si/CdTe Compton camera can be used efficiently below several hundred keV in particular.
Application of a stack of Si detectors to higher $\gamma$-ray energies,
where multiple Compton scattering has to be taken into account,
is described in \cite{ref:KamaeNIM1987, ref:KamaeIEEE1988, ref:TakahashiSPIE2003}.

\section{High resolution CdTe pixel detector and Double-sided Si Strip Detector}

\begin{figure}
\centerline{\includegraphics{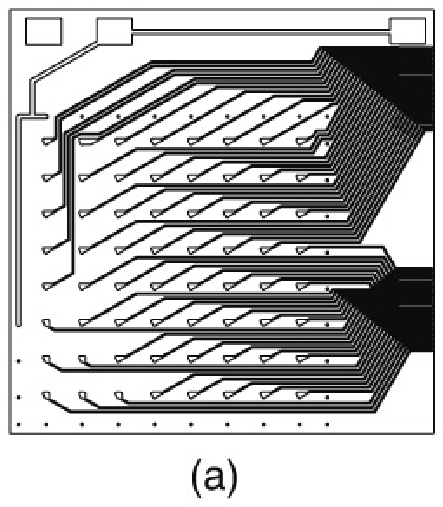}}
\vspace{5mm}
\centerline{\includegraphics{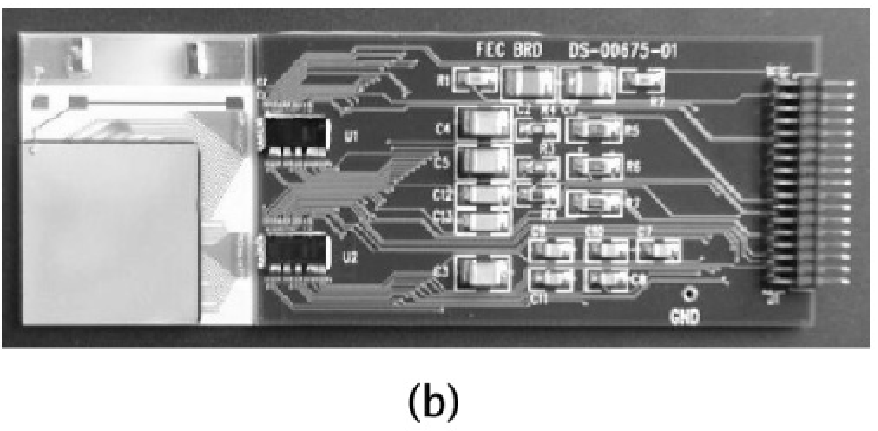}}
\caption{ (a) Layout of the fanout board for the CdTe pixel detector. It consists of bump
pads and patterns to route the signal from pads on the surface of the ceramic board,
which has a thickness of 200~$\mu$m.
(b) A CdTe 8$\times$8 pixel detector. The signal of each pixel is fed into VA32TAs on FEC.}
\label{fig:8x8}
\end{figure}

As a key component of a Si/CdTe Compton camera,
we have developed a CdTe pixel detector (Fig.~\ref{fig:8x8}(b)). 
It is made of a CdTe diode and
has the dimensions of 16.5~mm $\times$ 16.5~mm as an active area,
surrounded by a 1~mm wide guard-ring, and a thickness of 0.5~mm.
On the cathode side, it is pixellated with 2~mm square electrodes.
The gap between pixels is 50 $\mu$m.
A fanout board, which consists of bump pads
and patterns to route the signal from pads on the surface of the ceramic board,
 is shown in Fig.~\ref{fig:8x8}(a).
Once each pixel electrode on the CdTe diode is connected to the bump pad,
we can extract signals from the detector. 
The leakage current of this device is
200~pA/pixel at a bias voltage of 600~V and a temperature of 20\degree.
For the readout of many channels from the detector,
we utilize a newly developed low noise analog ASIC,
``VA32TA" \cite{ref:TajimaSPIE, ref:IEEE_VA}.
We handle 64 channels as a module and read out
with the Front End Card (FEC) on which two VA32TAs are mounted.

\begin{figure}
\centering
\includegraphics{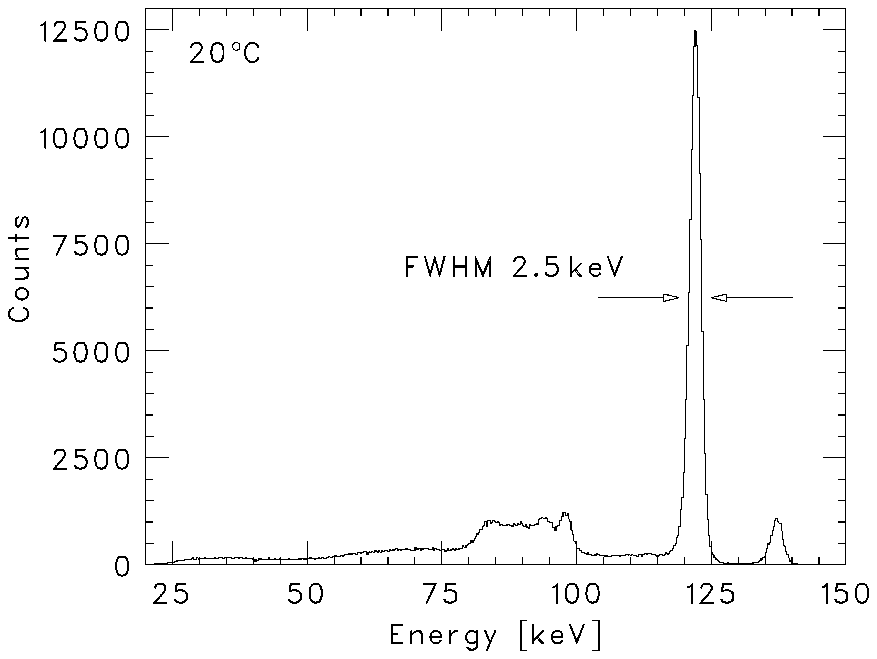}
\caption{Summed \co spectra of the CdTe pixel detector.
The operating temperature is 20\degree and the bias voltage is 600~V. 
The energy resolution is 2.5~keV. 
}
\label{fig:8x8spec}
\end{figure}

\begin{figure}
\centering
\includegraphics{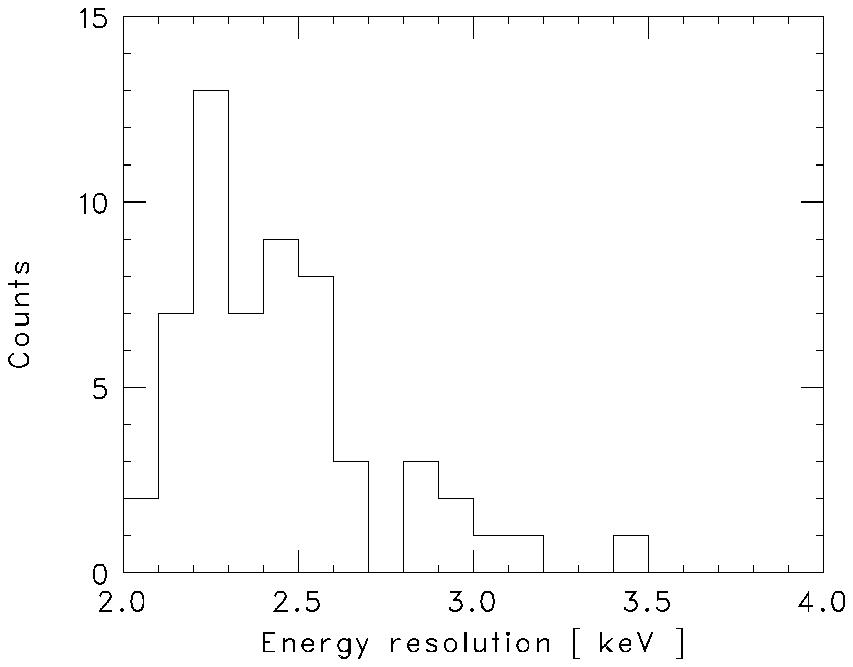}
\caption{Distribution of the energy resolution of each pixel of the CdTe detector.}
\label{fig:distres}
\end{figure}

We developed a first prototype detector and obtained spectra from
each channel. 
Fig.~\ref{fig:8x8spec} shows the sum of the energy spectra for all
channels except for seven channels at the corner in which no signal was obtained.
The detector is operated at room temperature (20\degree)
under the bias voltage of 600~V.
We obtain an
energy resolution of 2.5~keV (FWHM) for the 122~keV $\gamma$-ray lines.
Fig.~\ref{fig:distres} shows the distribution of the energy resolution of all channels.
It ranges from 2.1 to 3.4~keV.
When the operating temperature is 0\degree,
the highest energy resolution of
1.7~keV at 60~keV is obtained.

Although we have demonstrated that our bump method provides high yield
when we handle small pixel size\cite{ref:BonnIEEE2004, ref:TakahashiIEEE2001_pixel},
no signal was obtained for seven channels out of 64 of the first detector
due to a failure in the procedure of connecting
the pixels to the fanout board.
To improve the yield of the connections, we have fabricated
another batch of detectors in different conditions of bump bonding.
In this version, the yield of the connections has increased to 99.6\%,
but the energy resolution deteriorates to $\sim$ 6~keV
under the same operating conditions at 20\degree.
We are working to improve this performance.
Despite the relatively poor energy resolution,
we use detectors from the second batch in the following experiment
because they provide larger acceptance than the case for the first batch.

Another key component is the low noise
double-sided Si strip detectors(DSSDs).
The DSSD used in the present detector has a thickness of 300 $\mu$m and the strips have
a length of 2.56~cm on each side.
The strip pitch of DSSD is
800 $\mu$m and the strip gap
is 100$\mu$m.
The signals are also processed via the same FECs as the CdTe pixel detectors.
See Tajima et al. for details on DSSD developments \cite{ref:TajimaSPIE, ref:IEEE_VA}.

\section{Prototype of Si/CdTe Compton Camera and its Performance}

\subsection{Setup}

\begin{figure}
\centerline{\includegraphics{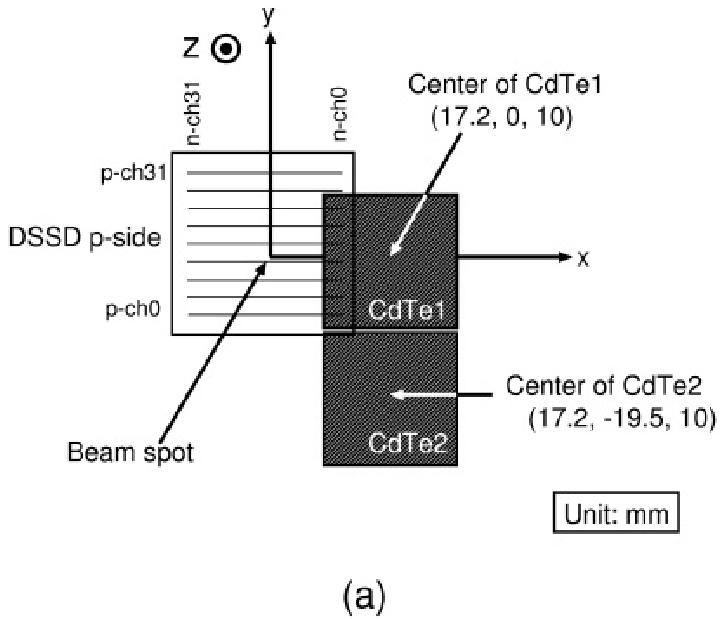}}
\centerline{\includegraphics{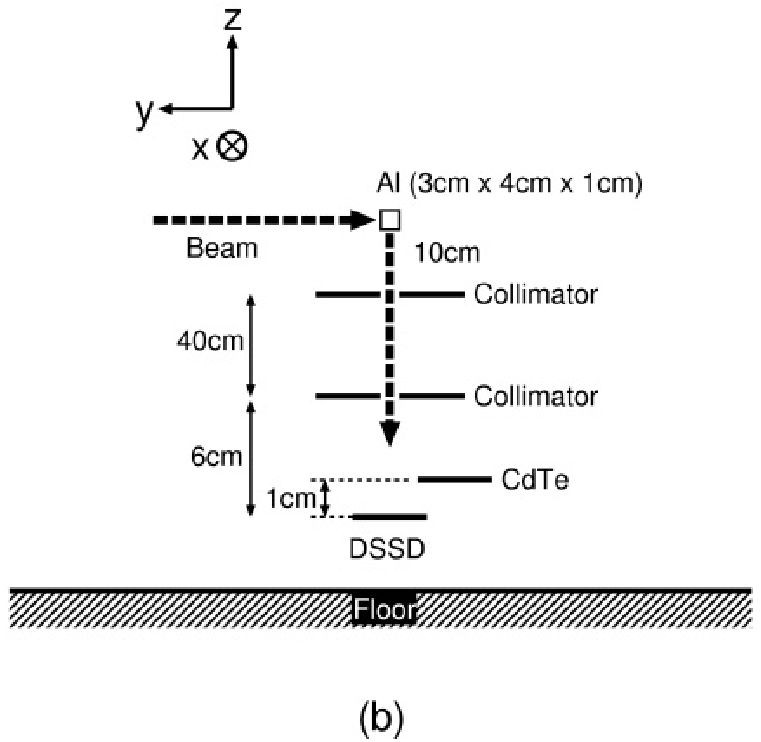}}
\caption{ (a) Conceptual image of the detector in top view.
A DSSD is put on the z~=~0~mm plane, and two CdTe pixel detectors
are put on the z~=~10~mm plane, where the beam spot on the DSSD is the origin. P-strips of the DSSD
are arranged to be parallel to the x-axis. The original beam
comes from $y=+\infty$.
(b) Horizontal view of the setup for the photon beam experiment.
To reduce the intensity, the original beam is scattered by an Al plate and collimated to
the prototype of a Compton camera. The photon beam enters the p-side of the DSSD.}
\label{fig:sp8setup}
\end{figure}

To demonstrate the concept of
the Si/CdTe Compton camera, 
two CdTe 8$\times$8 pixel detectors are combined with one DSSD.
In Fig.~\ref{fig:sp8setup}(a), we present a schematic view of the arrangement of the detectors.
In this configuration,
photons scattered in the DSSD are absorbed by the CdTe detectors
if the scattering angle is between
100$^{\circ}$ and 130$^{\circ}$.
This geometry is adopted because the scattered photons
have relatively low energy and can be photo-absorbed efficiently in CdTe.

The CdTe detectors used in the prototype
have an energy resolution of about 7~keV on average when we operate at
a temperature of 25\degree with a bias voltage of 500~V.
The energy resolution of the DSSD used in this system is 5~keV
at a temperature of 25\degree.
The resolution is deteriorated due to the rather high operation temperature.
As to the DSSD, ``RC chips'',
which provides bias voltage via
polysilicon bias resistors and AC-coupling between each strip and
the input of VA32TA, could also be the cause of
the degradation of the spectra.
Another type of DSSD we have developed with a strip pitch of 400 $\mu$m
has achieved an energy resolution of 
1.3~keV at a temperature of 0\degree without ``RC chips''
\cite{ref:TajimaSPIE, ref:IEEE_VA}.

Two CdTe pixel detectors and a DSSD are read out together by
the specially designed compact readout system including an ADC,
which is controlled
with a fast serial interface, ``Space Wire (IEEE 1355)''\cite{ref:IEEE_mitani}.
During data acquisition, events are taken when trigger signals are generated
from CdTe pixel detectors.
The data from triggered CdTe detectors and the DSSD are read out.

We conducted a photon beam experiment at
the synchrotron radiation facility, ``SPring-8'' \cite{ref:SPring8} in Japan on 7 July, 2003.
The initial beam energy was 270~keV. 
To reduce the intensity of the original beam, the beam was scattered by
an Al plate. Photons scattered at 90$^{\circ}$ are collimated with Pb collimators
as an input to the Compton camera (Fig.~\ref{fig:sp8setup}(b)).
According to the Compton kinematics, the estimated incident energy at
the detector is 177~keV
and the incident photons are nearly 100~\% linearly polarized.
The electric vectors of the photons are parallel to the p-strips of the DSSD.
We irradiated the photon beam with the geometry shown in Fig.~\ref{fig:sp8setup}(a).
After a irradiation, the whole Compton camera system is
rotated by 90$^{\circ}$ around the z-axis.
By performing this procedure four times, we obtain 360$^\circ$ coverage. 
During the experiment, the noise of n-strips of the DSSD was high and
we can not use the information from the n-strips.
We estimate the hit channel of the n-side on the basis of geometrical information
of the collimator and fix the hit channel in the following analysis.

\subsection{Reconstruction of Compton scattering angle}

\begin{figure}[th]
\centerline{\includegraphics{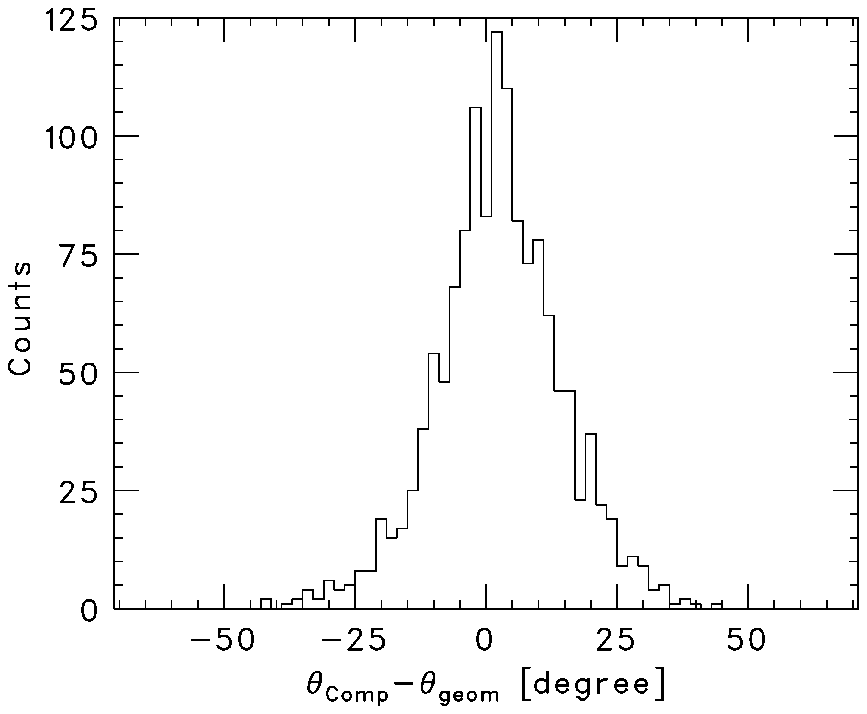}}
\caption{The distribution of $\theta_{\rm Comp} - \theta_{\rm geom}$
for the events under the conditions
150~keV$<$~$E_{\rm Si} + E_{\rm CdTe}<$180~keV,
$E_{\rm Si}>30$~keV and $90^{\circ}<\theta_{\rm Comp}<180^{\circ}$.
The angular resolution is calculated
to be 22$^{\circ}$(FWHM).}
\label{fig:exp_angres}
\end{figure}

From the data obtained, we reconstruct the Compton scattering angle.
Fig. \ref{fig:exp_angres} shows the distribution of $\theta_{\rm Comp} - \theta_{\rm geom}$
for the events under the conditions
150~keV$<$~$E_{\rm Si} + E_{\rm CdTe}<$180~keV,
$E_{\rm Si}>30$~keV and $90^{\circ}<\theta_{\rm Comp}<180^{\circ}$.
Here, $E_{\rm Si}$, $E_{\rm CdTe}$ are the energy deposited
in the DSSD and the CdTe detectors, respectively.
The scattering angle $\theta_{\rm Comp}$ is calculated from
Eq.~(\ref{eqn:comp}) using $E_{\rm Si}$, $E_{\rm CdTe}$ while
$\theta_{\rm geom}$ is that calculated from the hit positions under the assumption
that the incident beam is orthogonal to the DSSD surface.
The FWHM of the peak in Fig. \ref{fig:exp_angres}, i.e. the angular resolution of
the Compton camera, is 22$^{\circ}$. 
This value is consistent with
the angular resolution of 20$^{\circ}$ obtained from
an EGS4 Monte Carlo simulation\cite{ref:EGS4} with a low energy extension\cite{ref:EGS_KEK}.
From the analysis of the simulation,
the contribution to this angular resolution of the detector positional resolution is
estimated to be 6.2$^{\circ}$, that of the energy resolution 17$^{\circ}$, and
that of Doppler broadening 4.7$^{\circ}$.

Although the basic performance of a Si/CdTe Compton camera is demonstrated,
the angular resolution is not so pronounced.
If the energy resolution of both the DSSD and CdTe pixel detectors reaches their present best value,
1.3~keV and 2.5~keV, respectively, the contribution of the energy resolution can be
reduced to 3-4$^{\circ}$.
For the geometrical contribution, improvement of positional resolution
can be achieved either by using smaller pitch devices or by increasing the distance between
the DSSD and CdTe detectors.
Note that the relatively large Doppler broadening contribution
is only because of the low incident gamma-ray energy and 
there is no semiconductor currently, except for diamond, with better performance.

\begin{figure}[ht]
\centerline{\includegraphics{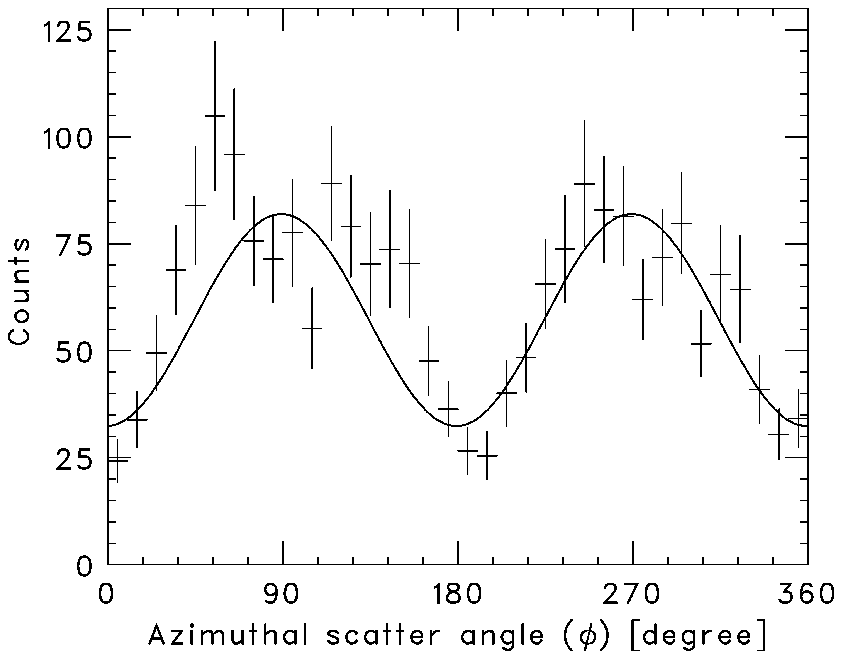}}
\caption{Experimental $\phi$ distribution corrected with a $\phi$ distribution obtained from
a simulation in which unpolarized photons are incident on the prototype.
The modulation factor is found to be 43\%.}
\label{fig:exp_mod}
\end{figure}

\begin{figure}[h]
\centerline{\includegraphics{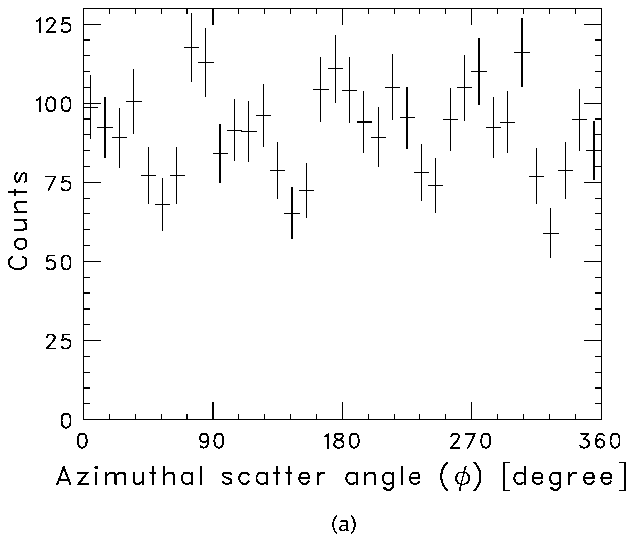}}
\centerline{\includegraphics{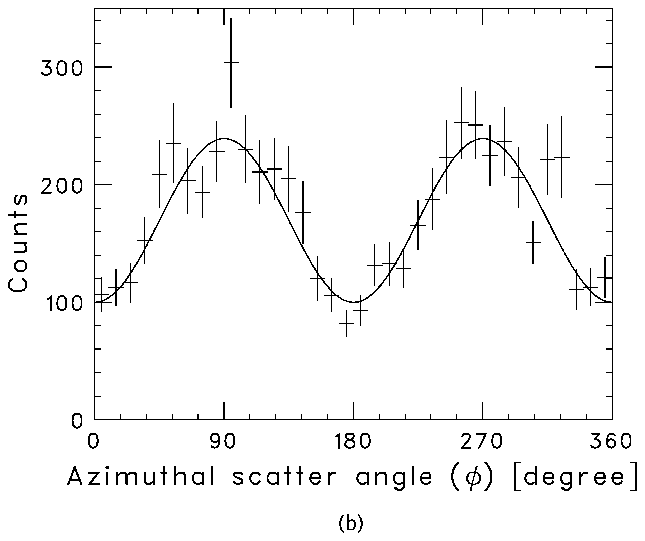}}
\caption{ (a) A $\phi$ distribution obtained from
an EGS4 Monte Carlo simulation in which unpolarized photons are incident on the prototype.
This shows the geometrical effect related to the layout of
the detectors and the associated quantization of
possible scattering angle. The statistical error bars arise from the limit of simulation time.
(b) A $\phi$ distribution obtained from an EGS4 Monte Carlo simulation where
a 100\% linear polarized beam is incident on the prototype.
The modulation factor is found to be 41\%.}
\label{fig:modulation}
\end{figure}

\subsection{Performance as a polarimeter}

An important feature of the Compton camera  is the capability of
measuring polarization\cite{ref:polarimetry}. 
The differential cross-section of Compton scattering is 
expressed as
\begin{equation}
     \frac{{\mathrm d} \sigma}{{\mathrm d} \Omega} =  \frac{r^2_e \beta^2}{2}\left( \beta^{-1}+\beta-2\sin^2\theta\cos^2\phi\right) .
     \label{eqn:pol}
\end{equation}
Here, $\theta$ is the scattering angle,
$\phi$ is the azimuthal angle of the scattered photon with respect to 
the electric vector of the incident photon, $\beta = E_2/(E_1+E_2)$ and $r_e$ is
the classical electron radius.
As shown in Eq.~(\ref{eqn:pol}), photons are likely to be scattered
in a direction orthogonal to their electric vector.
Therefore the $\phi$ distribution of scattered photons provides
information of  the linear polarization of incident photons.

The azimuthal scattering angle ($\phi$) distribution
of the experiment is shown in Fig.~\ref{fig:exp_mod} and
an angular modulation is clearly observed.
To correct the geometrical effect related to the layout of
the detectors and the associated quantization of
possible scattering angle, the measured
$\phi$ distribution
is normalized by
the model distribution (Fig.~\ref{fig:modulation}(a))
estimated from an EGS4 Monte Carlo simulation for an unpolarized beam.
Thus, Fig.~\ref{fig:exp_mod} includes the statistical error of both the
experiment and simulation.
For comparison,
a $\phi$ distribution obtained
from the results of a simulation with a 100\% linearly polarized beam
is shown in Fig.~\ref{fig:modulation}(b), which is corrected in the same way as the actual data.
In the simulation, 10$^8$ photons are incident on the detector and
6.3$\times$10$^2$ photons undergo one Compton scattering followed
by photoelectric absorption.
The low efficiency shown here must be improved by stacking Si and CdTe detectors.

We fit the distribution in Fig.~\ref{fig:exp_mod}, \ref{fig:modulation}(b)
with a formula,
\begin{equation}
N(\phi) = p_0 + p_1 \cos ^2 ( \phi + p_2 ) ,
\label{eqn:phidist}
\end{equation}
which is obtained by integrating Eq. (\ref{eqn:pol}) over the solid angle.
The polarimetric modulation factor, $\mathcal{Q}$,
\begin{equation}
  \mathcal{Q} \equiv \frac{N_{max}-N_{min}}{N_{max}+N_{min}} = \frac{p_1}{2p_0+p_1} ,
  \label{eqn:mod}
\end{equation}
is 43($\pm$3)\% for the experimental data,
which is consistent with that of the simulation data,
41($\pm$2)\%.

\section{Conclusion}

The hard X-ray and $\gamma$-ray bands are important frequency windows for
exploring the energetic universe. It is in these energy bands that
non-thermal emission becomes dominant.  In order to study the
``non-thermal universe", characterized by accelerated high energy
particles, we need an advanced mission which has orders of magnitude
better sensitivity than any previous or current missions.
To this end, we are working on a Si/CdTe Compton Camera.
With two CdTe pixel detectors and a DSSD,
we have constructed a prototype Si/CdTe Compton camera.
Using the prototype, we conducted a photon beam experiment
and demonstrate the concept of Si/CdTe Compton camera.
The angular resolution is 22$^{\circ}$ which
is limited by the energy resolution of 7~keV for the CdTe pixel detector
and 5~keV for the Si detector employed in this camera. 
By employing devices with a better energy resolution,
the angular resolution is expected to be improved significantly.
As the first demonstration of the Si/CdTe Compton camera
as a gamma-ray polarimeter, we have measured polarization
by using a synchrotron photon beam. We have succeeded to
obtain a very clear modulation pattern.
The polarimetric modulation factor is measured to be 43\%.
Based on these results,
a semiconductor Compton telescope utilizing
Si and CdTe is shown to posses a potential
as both a Compton camera and a polarimeter.

\section*{Acknowledgment}

The authors would like to thank P.~G.~Edwards and C.~Baluta for their critical
reading of the manuscript.




\end{document}